%% file: bare_conf.tex
\documentclass[10pt,conference]{IEEEtran}

\usepackage{graphicx}
\usepackage{cite}
\usepackage{url}
\usepackage{amsmath}
\usepackage{array}
\usepackage{enumitem}
\usepackage{caption}
\usepackage{subcaption}
\usepackage{hyperref}

\input{macro}

\title{The Ripple Effect of Vulnerabilities in Maven Central: Prevalence, Propagation, and Mitigation Challenges}

\author{
    \IEEEauthorblockN{Ehtisham Ul Haq}
    \IEEEauthorblockA{Lassonde School of Engineering\\ 
    York University, Toronto, Canada\\
    euh52@yorku.ca}
    \and
    \IEEEauthorblockN{Song Wang}
    \IEEEauthorblockA{Lassonde School of Engineering\\ 
    York University, Toronto, Canada\\
    wangsong@yorku.ca}
    \and
    \IEEEauthorblockN{Robert S. Allison}
    \IEEEauthorblockA{Lassonde School of Engineering\\ 
    York University, Toronto, Canada\\
    robert.allison@lassonde.yorku.ca}
}

\begin{document}

\maketitle

\begin{abstract}
The widespread use of package managers like Maven has accelerated software development but has also introduced significant security risks due to vulnerabilities in dependencies. In this study, we analyze the prevalence and impact of vulnerabilities within the Maven Central ecosystem, using Common Vulnerabilities and Exposures (CVE) data from OSV.dev and a subsample enriched with aggregated CVE data (CVE\_AGGREGATED), which captures both direct and transitive vulnerabilities. In our subsample of around 4 million releases, we found that while only about 1\% of releases have direct vulnerabilities, approximately 46.8\% are affected by transitive vulnerabilities. This highlights how a small number of vulnerable yet influential artifacts can impact a vast portion of the ecosystem. Moreover, our analysis shows that vulnerabilities propagate rapidly through dependency networks and that more central artifacts (those with a high number of dependents) are not necessarily less vulnerable. We also observed that the time taken to patch vulnerabilities, including those of high or critical severity, often spans several years. Additionally, we found that dependents of artifacts tend to prefer presumably non-vulnerable versions; however, some continue to use vulnerable versions, indicating challenges in adopting patched releases. These findings highlight the critical need for improved dependency management practices and timely vulnerability remediation to enhance the security of software ecosystems. 
\end{abstract}
\section{Introduction}

Extensive code reuse through package managers has become integral to modern software development, enabling rapid application building by leveraging existing libraries. Maven, a prominent package manager for Java, simplifies dependency management by automating the inclusion of both direct and transitive dependencies. However, this convenience introduces significant security risks. Vulnerabilities in dependencies can propagate through complex and deep dependency chains, potentially affecting a vast number of projects \cite{Pashchenko2018, harzevili2023characterizing}.

The problem is exacerbated by the fact that developers often lack visibility into the vulnerabilities present in their transitive dependencies. Studies have shown that a considerable number of software projects inadvertently include vulnerable libraries \cite{Ponta2018}. Although not all vulnerabilities necessarily impact dependent projects, as their exploitability depends on code usage, those that do have led to significant security breaches and data leaks. For instance, the Equifax data breach in 2017, which exposed the personal information of millions of individuals, was attributed to an unpatched vulnerability in a third-party library \cite{Spitzner2018}. This incident highlights the importance of timely vulnerability management in dependencies.

Despite awareness of these risks, many developers struggle to keep their dependencies updated. Kula et al. found that outdated libraries with known vulnerabilities are prevalent in software projects, mainly due to challenges in dependency management and the fear of breaking changes \cite{Kula2018}. Addressing these issues is essential for improving software security. By analyzing the prevalence and propagation of vulnerabilities, we can identify areas for improvement and develop strategies to mitigate risks associated with dependency management.

In this paper, we aim to fill the gap in understanding how vulnerabilities affect the Maven Central ecosystem. Specifically, we address the following research questions:

\begin{enumerate}[label=\textbf{RQ\arabic*:}, leftmargin=4em]
    \item (\textbf{Distribution}) What proportion of releases have known vulnerabilities? What is the proportion of releases directly and transitively impacted?
    \item (\textbf{Propagation}) How do vulnerabilities propagate through the dependency network, and which projects are most affected?
    \item (\textbf{Lifetime}) What is the average time taken to patch a vulnerability in a dependency?
    \item (\textbf{Response}) How do users of an artifact react to the discovery of a vulnerability in that artifact?
\end{enumerate}

To address these questions, we analyzed a dataset from the Maven Central repository, enriched with Common Vulnerabilities and Exposures (CVE) data sourced from OSV.dev \cite{OSV2023}. Using Neo4j and Cypher queries, we conducted experiments to uncover empirical insights into the prevalence and propagation of vulnerabilities, the timeliness of patches, and user behavior in response to vulnerabilities. 

\section{Experiments and Results} 
The experiments utilized the latest dataset snapshot from August 30th, 2024, comprising 658,078 artifact nodes, 14,459,139 release nodes, and 134,119,545 dependency edges. The database was enriched with CVE data up to September 4, 2024. We added additional CVE\_AGGREGATED values (capturing both direct and transitive vulnerabilities) using version 2.1.0 of the Goblin Weaver tool \cite{Goblin2025}. Due to the dataset's extensive size and the computational intensity of the enrichment process, the CVE\_AGGREGATED data enrichment was limited to a subset of 4,095,768 releases.

\subsection{RQ1: Distribution of Vulnerabilities}
\subsubsection*{\textbf{Experiment 1}}
We examined the subsample enriched with CVE\_AGGREGATED data to analyze the prevalence of vulnerabilities. Using Cypher queries on JSON-formatted CVE data, we identified releases with at least one direct vulnerability (non-empty CVE field) and those with only transitive vulnerabilities (non-empty CVE\_AGGREGATED but empty CVE field).

\begin{figure}[h]
    \centering
    \includegraphics[width=0.8\linewidth]{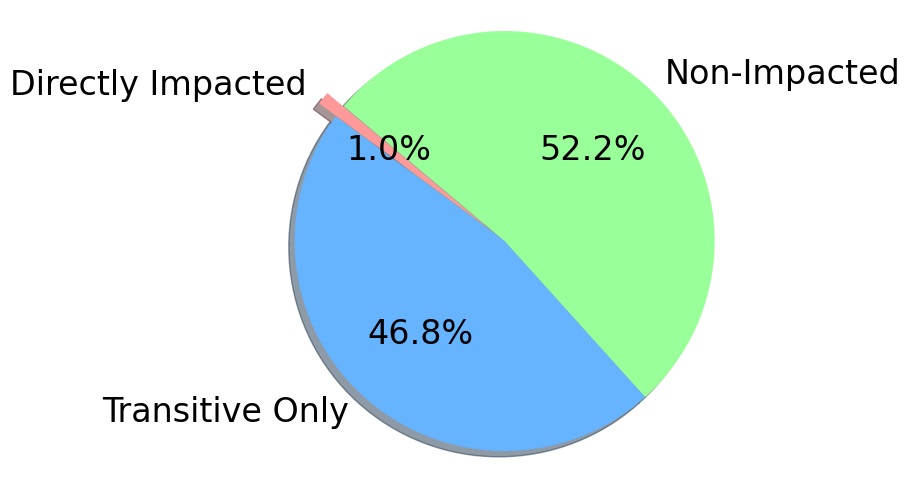}
    \caption{Proportion of Vulnerable Releases (from 4,095,768 Releases with Enriched CVE\_Aggregated Data)}
    \label{fig:1}
\end{figure}

\textbf{\textit{Result:}} We found that 40,809 releases (1\% of the subset) had at least one direct vulnerability and 1,916,314 releases (46.8\%) had at least one transitive vulnerability. Figure~\ref{fig:1} illustrates the proportion of releases affected by vulnerabilities.

\mybox{
\textbf{\textit{Finding 1:}} These results show that while direct vulnerabilities impact a very small fraction of releases, transitive vulnerabilities affect nearly half. It underscores the importance of considering both direct and transitive dependencies when assessing a project's security posture.
}

\subsection{RQ2: Propagation of Vulnerabilities}

\subsubsection*{\textbf{Experiment 2.1}}
To analyze how vulnerabilities propagate through the dependency network, we examined artifacts with direct vulnerabilities: 
\begin{itemize} \item From the complete dataset (around 14 million releases), we identified 77,393 releases with at least one direct vulnerability. \item These releases were linked to 1,411 distinct artifacts, which we labeled as \textit{ArtifactsWithDirectVulnerabilities}. \item Using Cypher queries, we calculated the number of dependent releases at: 
    \begin{itemize}
        \item Depth 1: Direct dependents of vulnerable artifacts.
        \item Depth 2: Dependents of direct dependents at depth 1.
    \end{itemize}
\end{itemize}

\begin{figure}[h]
    \centering
    \includegraphics[width=0.75\linewidth]{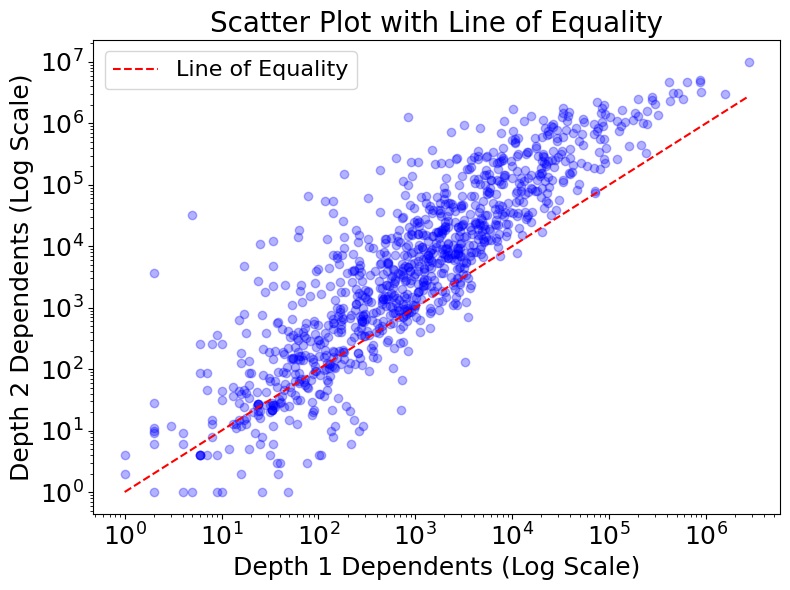}
    \caption{Depth 2 dependents vs. depth 1 dependents of \textit{ArtifactsWithDirectVulnerabilities} on a logarithmic scale}
    \label{fig:2}
\end{figure}

\textbf{\textit{Result:}} The average dependents per artifact were 17,993 at depth 1 and 142,948 at depth 2. Although 292 artifacts had no direct dependents, the remaining 1,119 showed significantly higher numbers at depth 2, as illustrated in Figure~\ref{fig:2}.

\mybox{
\textbf{\textit{Finding 2:}} The exponential growth in the number of dependents from depth 1 to depth 2 highlights the cascading effect of vulnerabilities in the dependency network. This vast amplification of impact underscores the need for strategies to manage and contain vulnerabilities at their source, as their unchecked propagation can affect a disproportionately large number of releases.}

\subsubsection*{\textbf{Experiment 2.2}}

To examine whether central artifacts (those with many dependents) are less vulnerable, we analyzed artifacts with aggregated vulnerabilities of moderate to critical severity. This included 1,953,496 releases from 145,174 artifacts labeled as \textit{ArtifactsWithSignificantVulnerabilities}. We calculated unique vulnerability counts to avoid over-penalizing artifacts for unpatched versions. The artifacts were grouped into four centrality buckets (\textit{Low}, \textit{Medium}, \textit{High}, \textit{Very High}) based on the number of dependents and the average vulnerabilities per bucket were calculated.

\begin{figure}
    \centering
    \includegraphics[width=0.75\linewidth]{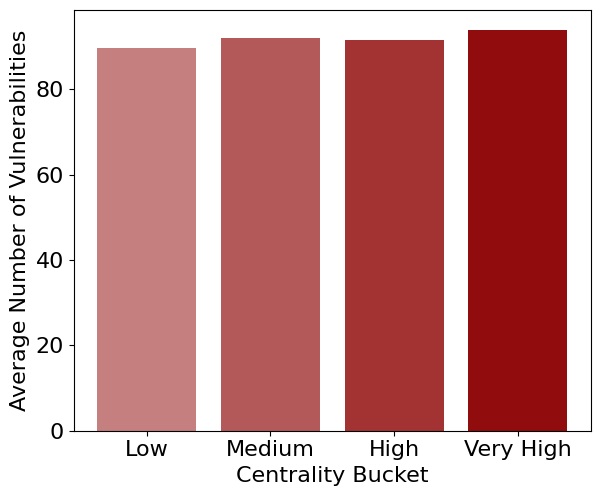}
    \caption{Average Vulnerabilities Across Centrality Buckets (Low to Very High)}
    \label{fig:3}
\end{figure}

\textbf{\textit{Result:}} The analysis revealed a subtle increase in vulnerabilities with centrality. Statistical tests confirmed significant differences across centrality buckets (ANOVA: F=4.26, p=0.005; Kruskal-Wallis: H=109.92, p$<$1e-22), with post-hoc analysis showing more vulnerabilities in the 'Very High' bucket compared to the 'Low' bucket (Tukey HSD: p=0.002). Figure~\ref{fig:3} shows the average vulnerabilities by centrality.

\mybox{ \textbf{\textit{Finding 3:}} The result suggests that centrality is not a safeguard against vulnerabilities. Although OWASP Dependency Check recommends widely adopted libraries for their perceived stability \cite{owasp_leverage_libraries}, our findings reveal that such libraries can still harbor significant risks. This highlights a dual responsibility: maintainers of central libraries must uphold rigorous security practices, and developers must assess dependencies for vulnerabilities. This is important to prevent amplified risks across downstream projects.}

\subsubsection*{\textbf{Experiment 2.3}}
To identify the most affected projects, we used the \textit{ArtifactsWithSignificantVulnerabilities} data along with the unique significant vulnerability counts from Experiment 2.2 to find the top 10 artifacts with the highest number of distinct significant vulnerabilities. We also plotted the distribution of distinct significant vulnerabilities across all artifacts in \textit{ArtifactsWithSignificantVulnerabilities}.

\begin{figure}[h]
    \centering
    \includegraphics[width=0.75\linewidth]{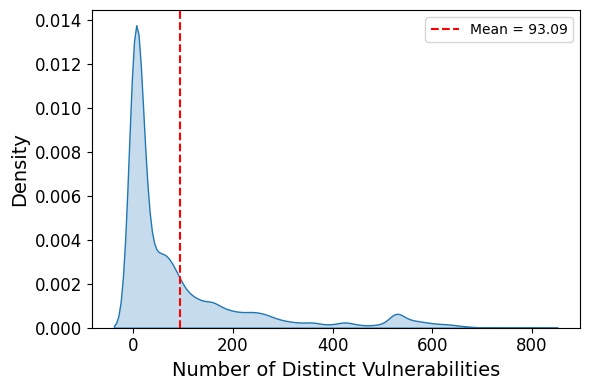}
    \caption{Kernel Density Estimation (KDE) of the Number of Distinct Vulnerabilities with Mean Line}
    \label{fig:4}
\end{figure}

\textbf{\textit{Results:}} The top 10 most affected artifacts had significant vulnerability counts ranging from 700 to 800, primarily belonging to the \textit{Apache Camel} and \textit{WildFly Camel} projects. Figure~\ref{fig:4} shows the distribution of distinct significant vulnerabilities.

\mybox{
\textbf{\textit{Finding 4:}} The skewed distribution of vulnerabilities indicates that they are concentrated in a small subset of high-impact artifacts, making them priority targets for security improvement. For example, \textit{Apache Camel} and \textit{WildFly Camel}, key integration frameworks facilitating communication between systems, have a high number of vulnerabilities. Vulnerabilities in such critical projects pose significant risks to the wide range of applications that rely on them, emphasizing the need for prompt remediation.} 

\subsection{RQ3: Lifetime of Vulnerabilities}

\subsubsection*{\textbf{Experiment 3}} To evaluate the timeliness of vulnerability patches, we analyzed the patch times for vulnerabilities in \textit{ArtifactsWithDirectVulnerabilities}. For each artifact-vulnerability pair, we recorded the earliest release timestamp where the vulnerability first appeared. Next, we identified the earliest subsequent release where the vulnerability was no longer present. If no such release was found, we marked the patch time as -1 to indicate an unpatched vulnerability. Finally, we grouped vulnerabilities by severity, identified their patch statuses, and calculated the respective patch times.

\subsubsection*{\textbf{Limitations}} The analysis does not account for vulnerabilities that reappear in later releases after being patched. Moreover, the analysis tracks vulnerability presence from its first recorded appearance, not its public disclosure date. Thus, patch times may include periods before disclosure, but they do reflect the actual exposure duration within the ecosystem.

\begin{figure}[ht]
    \centering
    
    % Subfigure 5a
    \begin{subfigure}{0.45\linewidth}
        \centering
        \includegraphics[width=\linewidth]{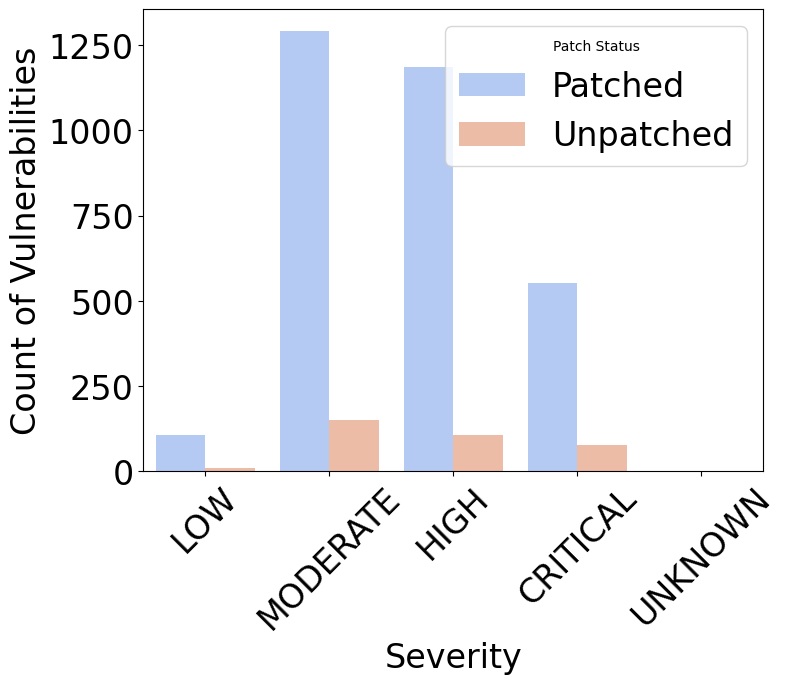}
        \caption{Patched vs. Unpatched Vulnerabilities by Severity}
        \label{fig:5a}
    \end{subfigure}
    \hfill
    % Subfigure 5b
    \begin{subfigure}{0.45\linewidth}
        \centering
        \includegraphics[width=\linewidth]{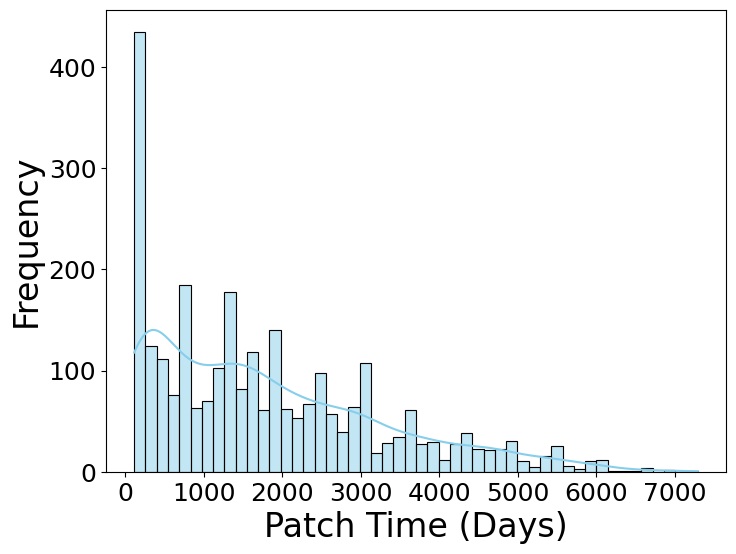}
        \caption{Distribution of Patch Times for Vulnerabilities}
        \label{fig:5b}
    \end{subfigure}
    \hfill
    % Subfigure 5c
    \begin{subfigure}{0.45\linewidth}
        \centering
        \includegraphics[width=\linewidth]{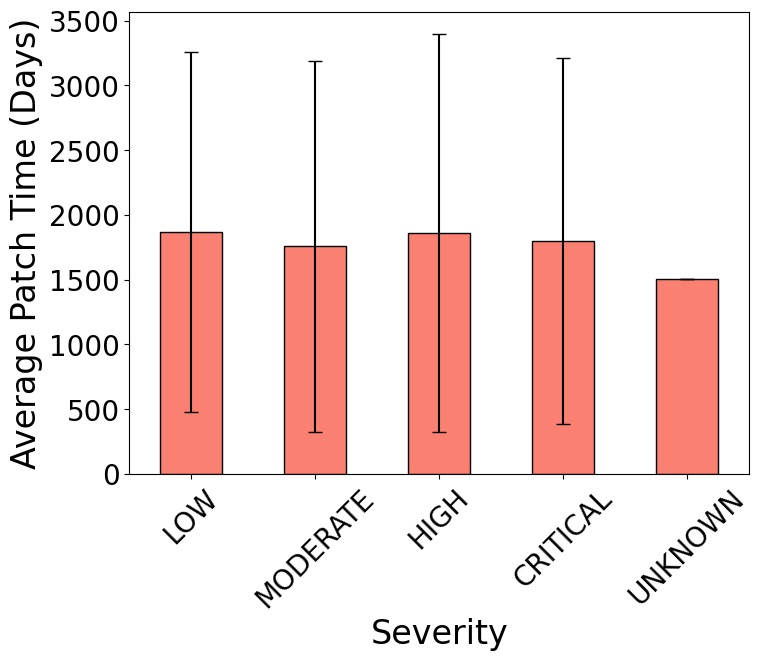}
        \caption{Average Patch Time by Severity ± Standard Deviation}
        \label{fig:5c}
    \end{subfigure}

    \caption{Vulnerabilities Patch Analysis}
    \label{fig:grouped}
\end{figure}

\textbf{\textit{Results:}} 
\begin{itemize} \item Patch Status by Severity: 
Figure~\ref{fig:5a} shows that while the majority of vulnerabilities were patched across all severity levels, a notable proportion remained unpatched. 'Low' severity vulnerabilities were relatively rare, and those with 'Unknown' severity were negligible.

\item Distribution of Patch Times: 
Figure~\ref{fig:5b} (histogram) shows most patches occurred within around 150 days. Then, the distribution tapers off.

\item Average Patch Time by Severity: 
Figure~\ref{fig:5c} shows average patch times of around 1,700 days across severities. All severities exhibit considerable variability (ranging from 300 to 3300 days) except 'unknown,' which shows almost none, likely due to its negligible sample size.
 
\end{itemize}

\mybox{
\textbf{\textit{Finding 5:}} These findings indicate that while many vulnerabilities are eventually patched, the time-frames are often extensive, leaving dependencies vulnerable for prolonged periods.}

\subsection{RQ4: Response to Vulnerabilities}

\subsubsection*{\textbf{Experiment 4}} 
We examined whether vulnerabilities influence the popularity of releases from an artifact, expecting users to favor non-vulnerable versions. 

\begin{itemize}
    \item Popularity Metric: We used the POPULARITY\_1\_YEAR metric enriched by Goblin Weaver \cite{Goblin2025}, representing the number of dependents of a release over a one-year window.
    \item Data Extraction: We extracted releases with popularity values greater than zero for \textit{ArtifactsWithDirectVulnerabilities}. \item Vulnerability Status: For every distinct vulnerability in an artifact, we determined whether each release was vulnerable or non-vulnerable based on its timestamp relative to the vulnerability’s first appearance and patch time. \item Analysis: We calculated the average popularity of vulnerable versus non-vulnerable releases for each artifact-vulnerability pair. 
\end{itemize}

\textbf{\textit{Results:}} We found that non-vulnerable releases had an average popularity advantage of approximately \textit{340} over vulnerable releases. Moreover, Figure 6 shows moving average trends, indicating that non-vulnerable releases tend to have significantly higher popularity than vulnerable ones of the same artifact.

\begin{figure}[h]
    \centering
    \includegraphics[width=0.75\linewidth]{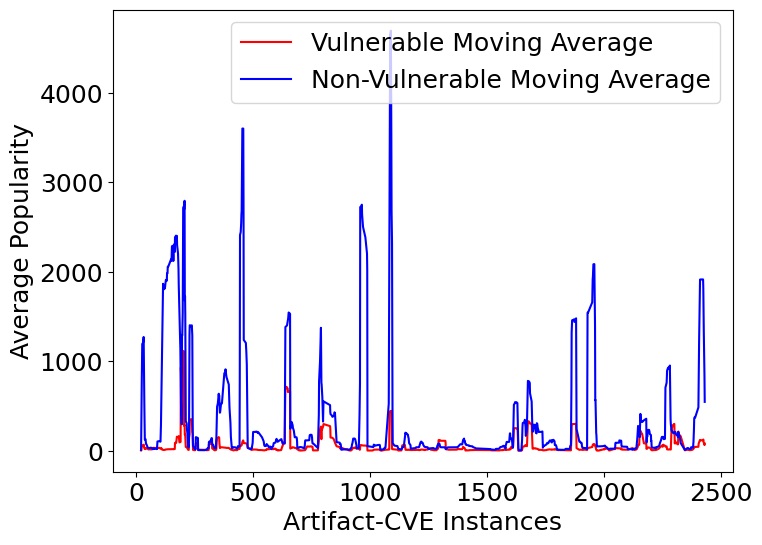}
    \caption{Moving Average Trend Lines Showing the Popularity of Vulnerable vs. Non-Vulnerable Releases for Each Distinct Artifact-Vulnerability Pair}
\end{figure}

\mybox{ \textbf{\textit{Finding 6:}} These results suggest that users favor releases without known vulnerabilities when possible. However, the continued use of vulnerable releases by some indicates factors like compatibility constraints, lack of awareness, limited alternatives, or inertia from prior integration of the package in existing systems, which may hinder the adoption of patched versions.}

\section{Related Work}
Vulnerabilities in software dependency networks have been widely studied across ecosystems, reflecting the challenges posed by their deep dependency chains.

Düsing and Hermann analyzed vulnerabilities across Maven, npm, and PyPI, highlighting slower patch adoption in Maven compared to npm \cite{Dusing2022}. Moreover, Kikas et al. explored the evolution of dependency graphs, revealing how increasing complexity introduces maintenance challenges and obscures vulnerabilities \cite{Kikas2017}. Similarly, Decan et al. analyzed the npm package network, observing a growing prevalence of transitive vulnerabilities and increasing delays in fixes, particularly for medium- and high-severity issues \cite{Decan2018}.

Our work builds upon these studies by quantifying vulnerability prevalence and propagation across dependency depths, analyzing patch timelines by severity, identifying risks in high-centrality artifacts, and exploring user behavior in response to vulnerabilities. These insights aim to inform and improve security practices in Maven, a central tool in Java development that powers countless enterprise and open-source projects.

\section{Conclusion}
In this study, we analyzed the prevalence and propagation of vulnerabilities in the Maven Central ecosystem using enriched CVE and CVE\_AGGREGATED data. Our findings revealed that while direct vulnerabilities impact only a small fraction of releases, transitive vulnerabilities affect nearly half, primarily due to dependencies on influential yet vulnerable artifacts. This underscores the need for enhanced security practices in critical projects to reduce their ecosystem-wide risk.

Vulnerabilities propagate exponentially at deeper dependency levels, highlighting the importance of addressing issues early to prevent widespread impact. Patch times for vulnerabilities, even high-severity ones, frequently span years, exposing systems to prolonged risks. Additionally, while users generally favor releases without known vulnerabilities, the continued reliance on vulnerable releases suggests challenges in adopting patched versions.

Future efforts should focus on developing efficient tools to detect and mitigate transitive vulnerabilities, such as enhanced dependency visualization and automated patch management solutions integrated into CI/CD pipelines. Encouraging the adoption of software composition analysis (SCA) tools and dependency health scoring can empower developers to make informed decisions. Promoting faster patching practices and addressing barriers to the adoption of secure releases through better documentation and awareness campaigns would further strengthen the ecosystem’s resilience.

\section{Data and Resources}

The Neo4j database dump used in this project, along with all the added labels, is available on Zenodo at \href{https://zenodo.org/records/14286455}{DOI: 10.5281/zenodo.14286455}. In addition, we provide supplementary materials, including a file containing the Cypher queries, Jupyter notebooks, and CSV files utilized for data analysis and visualizations.

\balance
\bibliographystyle{IEEEtran}
\bibliography{biblio}
\end{document}

%% file: macro.tex
\usepackage{soul}
\usepackage{hyperref}
\usepackage{multirow}
\usepackage{fancybox}
\usepackage{enumitem}
\usepackage{graphicx}
\usepackage{balance}
\usepackage{color}
\usepackage{float}
\usepackage{tabularx}
\usepackage{adjustbox}
\usepackage{xcolor}
\usepackage{array}
\usepackage{amsmath}
\usepackage{xspace}
\usepackage{url}
\usepackage{tikz}
\usepackage{caption}
\usepackage{pgfplots}
\usepackage{listings}
\usepackage{color}
\usepackage{graphicx}
\definecolor{dkgreen}{rgb}{0,0.6,0}
\definecolor{gray}{rgb}{0.5,0.5,0.5}
\definecolor{mauve}{rgb}{0.58,0,0.82}
\pgfplotsset{compat=1.16}
\usepackage{pgf-pie}
\usetikzlibrary{pgfplots.statistics,calc}
\usepackage{algorithm}
\usepackage{algpseudocode}
\usepackage{subcaption}
\usepackage{listings}

\usepackage{tcolorbox}
\makeatletter
\newcommand{\mybox}[1]{%
	\setbox0=\hbox{#1}%
	\setlength{\@tempdima}{\dimexpr\wd0+13pt}%
	\begin{tcolorbox}[boxrule=0.5pt, colback=white, arc=4pt,
		left=6pt,right=6pt,top=6pt,bottom=6pt,boxsep=0pt]
		#1
	\end{tcolorbox}
}

\definecolor{codegreen}{rgb}{0,0.6,0}
\definecolor{codegray}{rgb}{0.5,0.5,0.5}
\definecolor{codepurple}{rgb}{0.58,0,0.82}
\definecolor{backcolour}{rgb}{0.95,0.95,0.92}

\lstdefinestyle{mystyle}{
  language=Python,
  aboveskip=3mm,
  showstringspaces=false,
  columns=flexible,
  numbers=none,
  backgroundcolor=\color{backcolour},
  commentstyle=\color{codegreen},
 keywordstyle=\color{magenta},
    numberstyle=\tiny\color{codegray},
    stringstyle=\color{codepurple},
    basicstyle=\small\ttfamily,
    breakatwhitespace=false,         
    breaklines=false,                 
    captionpos=b,                    
    keepspaces=false,                 
    numbersep=5pt,                  
    showspaces=false,                
    showstringspaces=false,
    showtabs=false,                  
    tabsize=2,
    escapeinside=``
}
\definecolor{nima2}{RGB}{1.0, 0.49, 0.0}
\definecolor{songcolor}{RGB}{191,191,191}
\definecolor{reemcolor}{RGB}{0.9, 5.0, 0.8}
\definecolor{aruncolor}{RGB}{51,255,51}